\documentstyle[twocolumn,aps,prc,psfig]{revtex}
\begin{document}

\title{Isospin effects on two-particle correlation functions 
in E/A = 61 MeV $^{36}$Ar + $^{112,124}$Sn reactions}

\author{
R.~Ghetti$^{1}$, 
V.~Avdeichikov, 
B.~Jakobsson,
P.~Golubev\\
{\small \it Department of Physics, Lund University, Box 118, S-22100 Lund, Sweden}\\
J.~Helgesson \\
{\small \it School of Technology and Society, Malm\"o University, S-205 06 Malm\"o, Sweden}\\
N.~Colonna, 
G.~Tagliente \\
{\small \it INFN and Dip.\ di Fisica, V.~Amendola 173, I-70126 Bari, Italy}\\
H.W.~Wilschut,
S.~Kopecky, 
V.L.~Kravchuk \\ 
{\small \it Kernfysisch Versneller Instituut, Zernikelaan 25 NL 9747 AA Groningen, The Netherlands}\\
E.W.~Anderson, 
P.~Nadel-Turonski, 
L.~Westerberg \\ 
{\small \it The Svedberg Laboratory, Box 533, S-75121 Uppsala, Sweden}\\
V.~Bellini, 
M.L.~Sperduto,
C.~Sutera \\
{\small \it  Laboratori Nazionali del Sud (INFN), Via S.\ Sofia 44, I-95123 Catania, Italy}
}
\maketitle

\begin{abstract}
Small-angle, two-particle correlation functions  
have been measured for $^{36}$Ar+ $^{112,124}$Sn 
collisions at E/A = 61 MeV. 
Total momentum gated neutron-proton ($np$) 
and proton-proton ($pp$) correlations are 
stronger for the $^{124}$Sn-target. 
Some of the correlation functions for particle 
pairs involving deuterons or tritons 
($nd$, $pt$, and $nt$) also show a dependence on 
the isospin of the emitting source. 
\end{abstract}

\vspace{0.1cm}
\noindent
PACS number(s): 25.70.Pq

\vspace{0.1cm}
{\small 
\noindent
$^1$Corresponding Author: Roberta Ghetti, 
Department of Physics, Lund University, Sweden}
{\it E-mail address}: roberta.ghetti@nuclear.lu.se\\

\normalsize
The isospin dependence of the nuclear equation of state (EOS) 
is probably the most uncertain property of neutron-rich matter. 
This property is essential for the understanding of extremely 
asymmetric nuclei and nuclear matter as it may occur in the $r$-process 
of nucleosynthesis or in neutron stars \cite{Nova}. 
In order to study the isospin-dependent EOS, 
heavy ion collisions with isotope separated 
beam and/or target nuclei can be utilized \cite{Science}. 
In these collisions, excited systems are created 
with varying degree of proton-neutron asymmetry.  
A noticeable isospin dependence of the decay 
mechanism has been predicted 
\cite{Mueller,Colonna,LiPRL00,Ma,Liu}. 
Sensitive observables should be: 
pre-equilibrium neutron/proton emission ratio \cite{LiPRL97}, 
isospin ``fractionation'' \cite{Baran02,LiKo97,Xu,Tan}, 
isoscaling in multifragmentation \cite{Tsang}, 
neutron and proton flows \cite{LiPRC01}. 

Recently, the two-nucleon correlation function has been 
considered as a probe for the density dependence of 
the nuclear symmetry energy \cite{Chen,Chen2}. 
In this theoretical study with 
an isospin-dependent transport model (IBUU), 
it was shown that a stiff EOS causes high momentum 
neutrons and protons to be emitted almost simultaneously, 
thereby leading to strong correlations. 
A soft EOS delays proton emission, which weakens the 
$np$ correlation. In this paper we study experimental 
two-particle correlation functions of systems similar 
in size, but with different isospin. 
This work shows that, indeed, an isospin signal can 
be derived. 

Two-particle correlation functions 
were measured in E/A~=~61~MeV $^{36}$Ar-induced 
collisions on isotope-separated targets 
of $^{112}$Sn and $^{124}$Sn. 
The experiment was performed at the AGOR 
Superconducting Cyclotron of KVI (Groningen).  
The interferometer consisted of 16 CsI(Tl) 
detectors for light charged particles, mounted 
at a distance 56--66 cm from the target in the angular 
range 30$^{\rm o}$$\le$$\theta$$\le$114$^{\rm o}$, 
and 32 liquid scintillator neutron detectors, 
mounted 2.7~m from the target behind the ``holes'' 
of the CsI array, in matching positions  
to provide the $np$ interferometer \cite{NimTof}. 
In this analysis, only data from the angular 
range 60$^{\rm o}$$\le$$\theta$$\le$120$^{\rm o}$ 
were used for neutrons. Finally, 32 phoswich modules from 
the KVI Forward-Wall were mounted in the angular range 
6$^{\rm o}$$\le$$\theta$$\le$18$^{\rm o}$, 
to collect information on the centrality of the collision. 
At least one fragment in the Forward-Wall was always 
required in our selected events, which biases our 
data towards midperipheral collisions \cite{NPA00,NiAl}.
Energy thresholds for $p$, $d$, $t$ in the CsI(Tl) detectors 
were 8, 11, 14~MeV respectively, and for neutrons in the liquid 
scintillators 2.0~MeV. 
Details about the experimental setup and the particle energy 
determination are given in Refs.\ \cite{NimTof,NiAl,Volly}. 

\begin{figure}
\centerline{\psfig{file=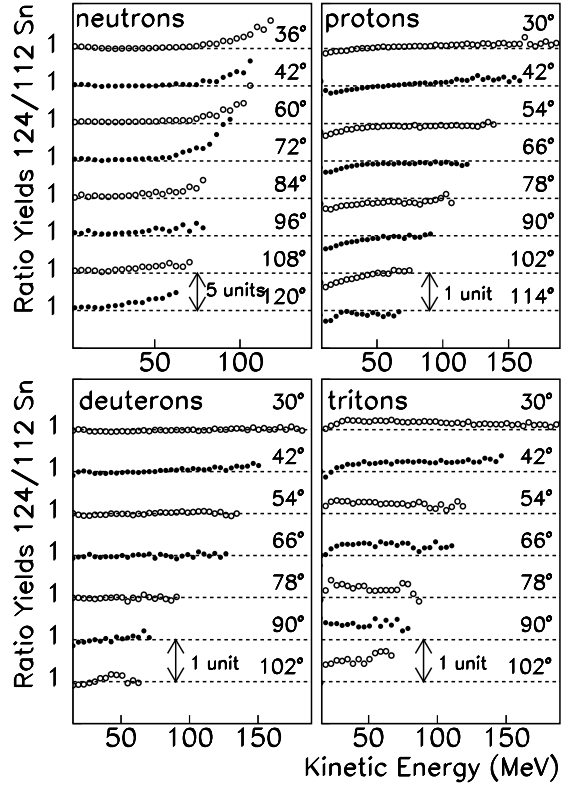,height=11.cm,angle=0}}
\caption{\small
Ratios of the $n$, $p$, $d$ and $t$ kinetic energy 
yields measured in $^{36}$Ar+$^{124}$Sn 
and $^{36}$Ar+$^{112}$Sn.
The ratios are arbitrarily shifted in the y-axis 
(by 5 units for $n$, by one unit for $p$, $d$, $t$). 
The dashed lines correspond to unitary ratios 
(at the angles indicated in the figure). 
}
\label{energy}
\end{figure}

Fig.\ \ref{energy} shows the ratios of the $n$, $p$, $d$ and 
$t$ kinetic energy yields measured in $^{36}$Ar+$^{124}$Sn 
and $^{36}$Ar+$^{112}$Sn (note the different scale 
in the figure for $n$ as compared to $p$, $d$ and $t$). 
An equal number of events is sorted for the two Sn-targets.  
The different solid angle coverage of $n$ and $p$ detectors 
is accounted for, and the neutron energy is efficiency 
corrected \cite{NimTof}. 
One can notice not only a substantial enhancement of the $n$  
yield for the neutron-rich system (as may be expected), 
but also that the enhancement is strongly energy dependent.
Furthermore, the $p$ yield is reduced at low energies 
for the neutron-rich system, and the $t$ yield is 
enhanced over the whole energy range. On the other hand, 
the yields of the deuteron spectra are the same for the 
two systems. 

The correlation function, 
$C(\vec{q},\vec{P}_{tot})$ = $k N_c(\vec{q},\vec{P}_{tot}) $
/ $N_{nc}(\vec{q},\vec{P}_{tot})$,
is constructed by dividing the coincidence yield ($N_c$) by the yield 
of uncorrelated events ($N_{nc}$) constructed from the product of the 
{\it singles} distributions \cite{NPA00}. 
$\vec{q} = \mu (\vec{p}_{1} / m_1 - \vec{p}_{2} / m_2)$, 
where $\mu$ is the reduced mass, is the relative momentum,
and  $\vec{P}_{tot} = \vec{p}_{1}+\vec{p}_{2}$ is the total momentum of the particle 
pair. The correlation function is normalized to unity at large values of $q$, 
80$<$$q$$<$120 MeV/c for $pp$ and $np$, and 
160$<$$q$$<$200 MeV/c for all other particle pairs.

\begin{figure}
\centerline{\psfig{file=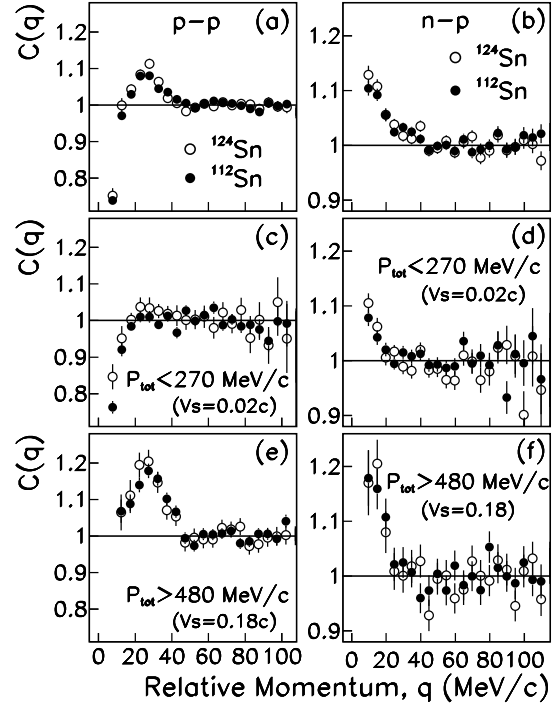,height=9.5cm,angle=0}}
\vspace{0.3cm}
\caption{\small
Angle-integrated (54$^{\rm o}$$\le$$\theta$$\le$114/120$^{\rm o}$) 
$pp$ (a,c,e) and $np$ (b,d,f) 
correlation functions from $^{36}$Ar+$^{112}$Sn (filled circles) 
and $^{36}$Ar+$^{124}$Sn (open circles).
Low-$P_{tot}$ (c,d) and 
high-$P_{tot}$ gated (e,f) correlation functions 
are also shown. 
}
\label{pp-np}
\end{figure}

The 54$^{\rm o}$$\le$$\theta$$\le$114/120$^{\rm o}$ 
$pp$/$np$ correlation functions are presented in 
Figs.\ \ref{pp-np}a,b. The neutron energy threshold 
is here 8 MeV, to match the proton threshold. 
The shape of the correlation functions looks as  
expected from the interplay of quantum statistical 
and final state interactions.
Comparing the two Sn-targets, one observes
a small but hardly significant 
enhancement of the correlation strength for $^{124}$Sn, 
in both $pp$ and $np$ correlation functions. 
For the interpretation of the correlation data, it is important 
to note that the correlation function depends on the space-time 
extent of the emitting source. From the size of the source, 
a stronger correlation is expected for the smaller 
$^{36}$Ar+$^{112}$Sn system, an effect expected also 
because of the larger excitation energy per particle available 
for this system (yielding a shorter emission time).
On the other hand, the change in neutron number implies a 
different symmetry energy which also affects the $n$ 
(and $p$) emission times. Neutrons are expected to be emitted 
faster in the neutron-rich system, which would lead to an 
enhancement of the correlation strength for $^{36}$Ar+$^{124}$Sn. 
Thus, the net influence on the correlation function 
is not easily predictable, both 
due to the uncertainty in the symmetry energy and to the 
presence of more than one source of emission. 

The emission of light particles from 61 A MeV 
(midperipheral) heavy ion reactions 
originates from (at least) three sources, a projectile-like (PLS) 
and a target-like (TLS) evaporative sources (statistical evaporation, SE) 
and an intermediate velocity source (IS). The IS source 
represents dynamical emission (DE), which is described either by early 
nucleon-nucleon collisions or by other pre-equilibrium processes. Some evidence 
for the creation of a ``neck-source'' with low density 
\cite{Montoya,Toke95,Laro97,Luka97,Pawl98,Laro99,Plag99,Mila00,InAl03} 
has recently been presented, and this may well comprise the 
dominating part of the DE source. 
By selecting nucleon pairs with high (low) total momentum, 
in the proper emission source frame, 
the DE (SE) emission should be favored. 
In the following, the $P_{tot}$ distributions 
for $pp$ and $np$ pairs are calculated in the frame of 
forward moving sources of velocity 0.02c (TLS) or 
0.18c (IS). These velocities are estimated from 
Maxwell-Boltzmann fits to the singles energy distributions. 
Low- and high-$P_{tot}$ selections 
correspond to 20--35$\%$ of the total correlation yields. 

Figs.\ \ref{pp-np}c,d present the 
$pp$ and $np$ correlation functions gated on low-$P_{tot}$. 
The $pp$ correlation function is sensitive 
to this gate, and a suppression in the 
correlation strength is observed, while the 
$np$ correlation function is only slightly suppressed.
This is a first hint that the $np$ correlation function 
may be dominated by DE. 

Figs.\ \ref{pp-np}e,f present the 
$pp$ and $np$ correlation functions gated on high-$P_{tot}$. 
Both $pp$ and $np$ correlation functions are 
affected and, in accordance with earlier observations \cite{NPA00}, 
an enhancement in the correlation strength is observed 
relative to the ungated correlation functions. 
\begin{figure}
\centerline{\psfig{file=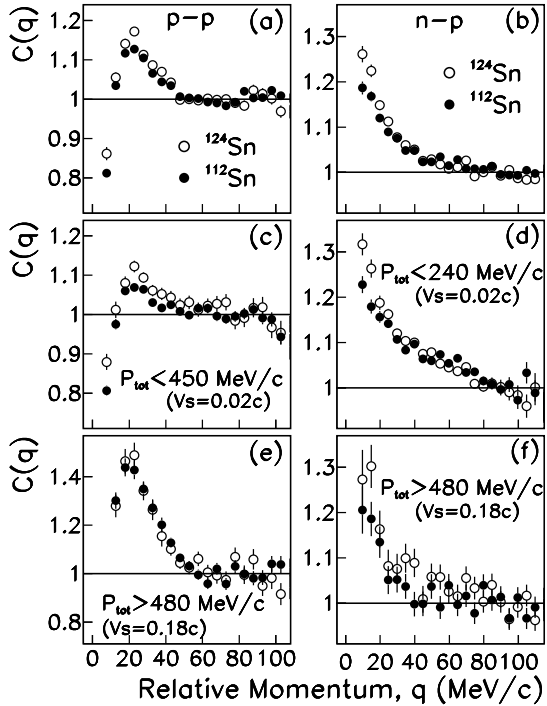,height=9.5cm,angle=0}}
\vspace{0.3cm}
\caption{\small
$pp$ correlation functions (a,c,e) 
measured in the forward detectors 30$^{\rm o}$$\le$$\theta$$\le$42$^{\rm o}$, 
and $np$ correlation functions (b,d,f) 
measured in the range 54$^{\rm o}$$\le$$\theta$$\le$120$^{\rm o}$, 
with $E_n$$\ge$2 MeV and $E_p$$\ge$8 MeV, from 
$^{36}$Ar+$^{112}$Sn (filled circles) 
and $^{36}$Ar+$^{124}$Sn (open circles).
Low-$P_{tot}$ (c,d) and 
high-$P_{tot}$ gated (e,f) 
correlation functions are also shown.
}
\label{fig-iso}
\end{figure}
All results from these angle integrated data 
(54$^{\rm o}$--114/120$^{\rm o}$) 
show smaller isospin effects
than could be expected from the $N/Z$ ratio, which 
is 1.18 for $^{36}$Ar+$^{112}$Sn and 
1.35 for $^{36}$Ar+$^{124}$Sn. 
In case the neck formation dominates the DE process, this is surprising, 
since the neck region should exhibit a strengthened neutron excess, quite 
sensitive to the isospin of the target nucleus \cite{Colonna}.
In an attempt to select phase-space regions that favor DE 
emission, we now look at: 
$i$) $pp$ correlations in the forward 
(30$^{\rm o}$$\le$$\theta$$\le$42$^{\rm o}$) region and 
$ii$)  $np$ correlations which include very low energy 
neutrons ($E_n$$\ge$2 MeV, in the lab system). Neutrons with 
2--8 MeV energy, detected at 54$^{\rm o}$$\le$$\theta$$\le$120$^{\rm o}$, 
correspond to 15-45 MeV energy in an IS ($v$=0.18c) source.

The results for $pp$ and $np$ correlation functions obtained 
under conditions  ($i$) and ($ii$), are shown in Fig.\ \ref{fig-iso}. 
A stronger $pp$ correlation function is now observed for the 
more neutron-rich system, $^{36}$Ar+$^{124}$Sn. 
This is true for integrated (a) as well as 
low-$P_{tot}$ gated (c) correlations. 
The low-$P_{tot}$ gated correlations show the 
largest isospin signal. 
The reason for the relatively large value of the $P_{tot}$ cut (450 MeV/c), 
is the necessity to populate the large $q$ (normalizing) region in 
the forward (30$^{\rm o}$$\le$$\theta$$\le$42$^{\rm o}$) detectors. 
This cut corresponds to TLS protons with $E_p$$<$30 MeV, and 
IS protons with $E_p$$<$20 MeV. It is these low energy protons 
that show the largest isospin effect.

The $np$ correlation functions at angles 54$^{\rm o}$--120$^{\rm o}$ 
(Figs.\ \ref{fig-iso}b,d,f), show large isospin 
effects when low energy neutrons, 2$\le$$E_n$$\le$8 MeV, 
are included (compare with Figs.\ \ref{pp-np}b,d,f). 
Again, the correlation is stronger for the 
more neutron-rich system, indicating that the space-time 
extent of the emitting source 
is smaller in $^{124}$Sn.
In particular, both low-$P_{tot}$ gated 
(here $P_{tot}$$<$ 240 MeV/c) 
correlation functions (Fig.\ \ref{fig-iso}d) are enhanced 
as compared to Fig.\ \ref{fig-iso}b 
(instead of being suppressed as expected for an 
evaporative source). 
A plausible explanation for this 
is that the pairs selected by this gate are dominated 
by DE particles. Since this effect did not occur 
in the data where the low energy neutrons were removed 
(Fig.\ \ref{pp-np}d), we attribute the correlation 
enhancement to DE particles in the energy range 15--45 MeV, 
which are favored in the neutron-rich system. 
If this isospin effect mainly comes from the intermediate energy 
particles, it should be connected to the density dependence 
of the nuclear symmetry energy \cite{Chen,Chen2}. 

\begin{figure}
\centerline{\psfig{file=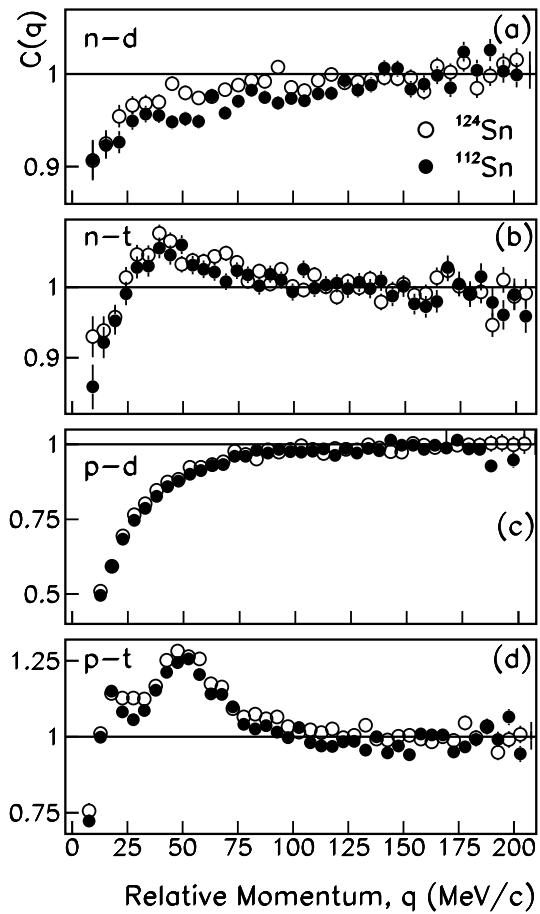,height=12.5cm,angle=0}}
\caption{\small
From $^{36}$Ar+$^{112}$Sn (filled symbols) 
and $^{36}$Ar+$^{124}$Sn (open symbols):
(a,b) $nd$, $nt$ 
(54$^{\rm o}$$\le$$\theta$$\le$120$^{\rm o}$); 
(c,d) $pd$, $pt$
(30$^{\rm o}$$\le$$\theta$$\le$42$^{\rm o}$).
}
\label{npdt}
\end{figure}

Within the multi-source reaction mechanism described above, 
composite particles, like deuterons and tritons, are believed to be 
predominantly emitted from the DE source \cite{Lanzano,Lefort}, 
where they are formed by a coalescence mechanism \cite{Coalescence}. 
Let us first remark that neither 
integrated nor $P_{tot}$ gated 
$dd$, $tt$ and $dt$ correlation functions 
show any appreciable difference between the 
two Sn-targets  \cite{Future}. This is  
in agreement with the small sensitivity shown by 
the calculations of Ref.\ \cite{Chen3}. 
Even so, a variation of the correlation functions 
such as $nd$, $nt$, etc.\ may be expected, 
as a consequence of the isospin effects on 
neutrons and protons. 
Indeed, this is the case in our experimental data, 
which can be taken as a further evidence for the presence 
of true isospin effects. 
Figs.\ \ref{npdt}a,b present the angle-integrated 
(54$^{\rm o}$$\le$$\theta$$\le$120$^{\rm o}$) 
$nd$ (a) and $nt$ (b) 
correlation functions, 
measured for the two Sn-targets, with a neutron energy 
threshold of 2 MeV. 
The anticorrelation observed for $nd$ pairs (Fig.\ \ref{npdt}a), 
has been observed earlier for smaller collision systems, 
both at lower energies \cite{Kryger} and for 61 A MeV \cite{NiAl}. 
It may originate from the depletion 
of low relative momentum $nd$ pairs due to triton 
formation \cite{Tomio}.
In Fig.\ \ref{npdt}a the difference in correlation 
strength between the two Sn-targets 
can be observed. 

The correlation functions of $nt$ pairs (Fig.\ \ref{npdt}b), 
exhibit a broad peak which contains the contributions from 
the particle-unbound ground state of $^4$H,  
and possibly from higher lying excited states \cite{Tilley}.
Once again, a small enhancement in the strength is observed 
for the $^{124}$Sn-target. 

The $pd$ and $pt$ correlation functions, measured in the forward 
angular range (30$^{\rm o}$$\le$$\theta$$\le$42$^{\rm o}$), 
are shown in Figs.\ \ref{npdt}c,d.
The $pd$ correlation function (c) is characterized by a pronounced 
anticorrelation at small $q$, due to final state Coulomb 
repulsion. The isospin effect is negligible. 
The $pt$ correlation function (d),
contains resonance contributions from several 
excited states of $^4$He \cite{Tilley}.
A small isospin effect is seen in this 
correlation function. 

In summary, isospin effects 
have been investigated in the 
E/A = 61 MeV $^{36}$Ar+$^{112}$Sn, $^{124}$Sn reactions, 
and, for the first time, correlation functions 
from systems similar in size but with different isospin 
have been experimentally determined. 
Both angle-integrated and total-momentum gated 
correlation functions for all different  
pairs of particles containing $n$, $p$, $d$ and $t$, 
have been measured. 
The largest effects from the isospin of the emitting system 
are seen in the $np$ correlation function.
In particular, gated $np$ correlation functions 
which should favor a dynamical emission source, 
show a stronger correlation for $^{124}$Sn than $^{112}$Sn. 
This could be explained by different time distributions, 
with a shorter average emission time for the neutron-rich 
system. Smaller isospin effects are also seen 
in $pp$, $pt$, $nd$ and $nt$ correlation functions. 
These experimental results demonstrate that two-particle 
correlation functions indeed provide an additional 
observable to probe the isospin dependence of the 
nuclear EOS.
\\

The authors wish to thank 
S.\ Brandenburg and the AGOR crew, 
F.\ Hanappe and the DEMON Collaboration.
This work was partly supported by the European Commission  
(Transnational Access Program, contract HPRI-CT-1999-00109), 
and by the Swedish Research Council (grant nr F 620-149-2001). 


\end{document}